\documentclass[]{pasj00}
\draft

\begin{document}
\SetRunningHead{K. Kohno et al.}{Molecular Gas in the Post-Starburst Galaxy NGC 5195}
\Received{2000/09/07}%{yyyy/mm/dd}
\Accepted{2002/05/12}%{yyyy/mm/dd}

\title{Diffuse and Gravitationally Stable Molecular Gas in the Post-Starburst Galaxy NGC 5195}

\author{Kotaro \textsc{Kohno}}
\affil{\it Institute of Astronomy, The University of Tokyo, 2-21-1, Osawa, Mitaka, Tokyo 181-8588}
\email{kkohno@ioa.s.u-tokyo.ac.jp}

\author{Tomoka \textsc{Tosaki}}
\affil{\it Gunma Astronomical Observatory, Nakayama, Takayama, Agatsuma, Gunma 377-0702}
\email{tomoka@astron.pref.gunma.jp}

\author{Satoki \textsc{Matsushita}}
\affil{\it Submillimeter Array, Harvard-Smithsonian Center for Astrophysics,\\ P.O. Box 824, Hilo, HI 96721-0824, U.S.A.} 
\email{smatsu@sma.hawaii.edu}

\author{Baltasar \textsc{Vila-Vilar\'o}}
\affil{\it Steward Observatory, The University of Arizona, Tucson, AZ 85721, U.S.A.}
\email{bvila@as.arizona.edu}

\author{Toshihito \textsc{Shibatsuka}}
\affil{Nobeyama Radio Observatory, Minamimaki, Minamisaku, Nagano 384-1305}
\affil{Department of Astronomy, The University of Tokyo, 7-3-1 Hongo, Bunkyo-ku, Tokyo 113-0033}
\email{shiba@nro.nao.ac.jp}

\and

\author{Ryohei \textsc{Kawabe}}
\affil{National Astronomical Observatory, 2-21-1, Osawa, Mitaka, Tokyo 181-8588} 
\email{kawabe@nro.nao.ac.jp}


\maketitle

\begin{abstract}
The Nobeyama Millimeter Array (NMA) has been used to make
aperture synthesis CO(1$-$0) observations of the post-starburst
galaxy NGC 5195.
CO(1$-$0) and HCN(1$-$0) observations 
of NGC 5195 using the Nobeyama 45 m telescope
are also presented.
High-resolution (\timeform{1".9} $\times$ \timeform{1".8}
or 86 pc $\times$ 81 pc resolution at $D$ = 9.3 Mpc) 
NMA maps show a strong concentration of CO emission 
toward the central a few $\times$ 100 pc region of NGC 5195,
despite the fact that the current massive star formation
is suppressed there.
The face-on gas surface density, $\Sigma_{\rm gas}$,
within the $r<2''$ or 90 pc region reaches $3.7 \times 10^3$ 
$M_\odot$ pc$^{-2}$
if a Galactic $N_{\rm H_2}/I_{\rm CO}$ conversion factor is applied.
The extent of the central CO peak is about $5''$, or 230 pc, and is
elongated along the E--W direction with two-armed spiral-like structures,
which are typical for barred disk galaxies.
The HCN-to-CO integrated intensity ratio on the brightness temperature
scale, $R_{\rm HCN/CO}$, is about 0.02 within the
central $r < 400$ pc region. This $R_{\rm HCN/CO}$
is smaller than those in starburst regions by a factor of $5-15$.
These molecular-gas properties would explain 
why NGC 5195 is in a post-starburst phase; 
most of the {\it dense} molecular cores
(i.e., the very sites of massive star formation) have been
consumed away by a past starburst event,
and therefore a burst of massive star formation can no longer last,
although a large amount of {\it low density} gas still exists.
We find a steep rise of the rotation velocity toward the center of NGC 5195.
As a consequence, the critical gas surface density for a local gravitational
instability of the gas disk becomes very high
($\Sigma_{\rm crit} \sim 6.9 \times 10^3$ $M_\odot$ pc$^{-2}$), 
suggesting that the molecular gas
in the central region of NGC 5195 is gravitationally {\it stable},
in contrast to that of starburst galaxies. 
We propose that dense molecular gas can not be formed from remaining diffuse
molecular gas because the molecular gas in the center
of NGC 5195 is {\it too stable} to form dense cores via gravitational 
instabilities of diffuse molecular gas.
The deduced very high threshold density seems to be due to 
a high mass concentration in NGC 5195.
The known trends on the occurrence and luminosity of nuclear star formation
in early-type galaxies can be understood naturally
if the high threshold density is characteristic for early-type galaxies.
\end{abstract}

\section{Introduction}

Recent studies on the molecular-gas properties in galaxies
have revealed an intimate relationship between starbursts and 
high-density ($n_{\rm H_2} > 10^4$ cm$^{-3}$) molecular gas; 
quantitative correlations between HCN(1$-$0) emission, 
which is an indicator of dense molecular gas contents,
and star-formation tracers, such as a far-infrared (FIR) continuum
and radio recombination lines, have been demonstrated 
(e.g., Solomon et al.\ 1992; Zhao et al.\ 1996; Paglione et al.\ 1997).
High-resolution images of HCN(1$-$0) emission with millimeter-wave 
arrays also show spatial coincidence of dense molecular material 
with the massive star-forming regions in starburst/star-forming galaxies
(e.g., Paglione et al.\ 1995; Kohno et al.\ 1999, but see Aalto et al.\ 2001).

Given the important role of dense molecular matter on massive star-
formation, it would be then important to investigate how they are formed.
A nonaxisymmetric distortion of the underlying potential in galaxies caused
by various dynamical effects, such as disk instability and
galaxy--galaxy interactions (e.g., Noguchi 1996),
is often invoked as an efficient 
mechanism which stuff interstellar matter (ISM) on the large
scale (from a few to a few $\times$ 10 kpc) disk of galaxies 
into the very central (a few $\times$ 100 pc) regions by removing the angular 
momentum of ISM (e.g., Shlosman et al.\ 1989).
To date, however, the physical process 
which could transform {\it diffuse} ISM into {dense} molecular
gas has not been addressed so often. Moreover, most of the
observational studies on dense molecular gas have concentrated on
luminous starburst galaxies, such as NGC 253 and M 82. 
It must be valuable to assess the distribution and
physical properties of molecular gas in {\it quiescent} and 
{\it less star-forming} galaxies
in order to seek a clue on dense gas formation. 

In this paper, we report on high-resolution aperture synthesis
CO(1$-$0) observations and simultaneous CO(1$-$0) as well as HCN(1$-$0)
spectroscopy of the early-type barred galaxy NGC 5195,
made with the Nobeyama Millimeter Array (NMA) and NRO 45 m telescope.
As described below, NGC 5195 may be one of the ideal targets
to investigate the molecular-gas properties in the galaxy whose
star formation is currently at a very low level;
NGC 5195 is a barred lenticular galaxy (SB0 pec; Sandage, Tammann 1981)
at a distance of 9.3 Mpc (Tully 1988), known as a companion of 
the well-studied spiral galaxy M 51/NGC 5194.
In spite of the abundant molecular
gas in the central region of NGC 5195 
(Sage, Wrobel 1989; Sage 1989, 1990; Aalto, Rydbeck 2001), the current
massive star-formation seems to be suppressed there. 
Narrow-band imaging does not show any significant H$\alpha$ {\it emission},
and strong H$\alpha$ {\it absorption} dominates in the central 
$\sim 10''$ region instead (Thronson et al.\ 1991; Sauvage et al.\ 1996; 
Greenawalt et al.\ 1998; 
see also optical spectra by Filippenko, Sargent 1985). 
This Balmer absorption feature is
attributed by a presence of numerous A-type stars, and 
few stars earlier than A4 exist in this region
(Rieke 1988; Yamada, Tomita 1996).
Considering the lifetime of A-type stars 
(from a few $\times$ $10^8$ to $10^9$ yr)
and the typical duration of starbursts 
(a few $\times$ $10^7$; Thornley et al.\ 2000
and references therein), it is strongly suggested that NGC 5195 
experienced a nuclear starburst about $\sim$ 1 Gyr before, 
and is now in the {\it post-starburst phase}, where the OB stars 
produced by the starburst event have disappeared and 
only late-type stars (A stars and later) remain in this region.
This situation is strikingly similar to the archetypical post-starburst
galaxy NGC 4736 (Pritchet 1977; Rieke et al.\ 1988; Walker et al.\ 1988;
Taniguchi et al.\ 1996).
A mid-infrared (MIR) study with ISOCAM by Boulade et al. (1996)
has also revealed the post-starburst nature of NGC 5195. 
Consequently, the current star-formation rate (SFR) is very small;
the H$\alpha$ emission luminosity can be found by subtracting
the underlying absorption feature by A stars, and  
is estimated to be $8.7\times10^{37}$
erg s$^{-1}$ within a $2'' \times 4''$ aperture (Ho et al.\ 1997a).
This is near the lowest end of $L$(H$\alpha$) in their sample,
containing about 500 nearby galaxies in the northern hemisphere.
Note that optical spectroscopy suggests the nucleus of NGC 5195
may be a LINER (e.g., Ho et al.\ 1997a), which could be a signature
of a low-luminosity active galactic nucleus (AGN).
Radio and MIR observations also show a compact and luminous
source in the center of NGC 5195 (van der Hulst et al.\ 1988;
Boulade et al.\ 1996).
It is, however, unclear whether a compact non-stellar
powered source exists in the nucleus of NGC 5195, because
neither a compact UV core (Barth et al.\ 1998) nor ionized
neon lines in the MIR band (Boulade et al.\ 1996),
both are characteristic features of an AGN, has been detected.
High-angular resolution X-ray observations of NGC 5195 also reveal an extended
distribution of hot gas (Ehle et al.\ 1995; Georgantopoulos et al.\ 2002), 
in favour of a star-forming origin for the 
bulk of the X-ray emission.
The radio and MIR concentrations, therefore, would be 
related to past starburst events (SNRs and
hot dust heated by late-type stars).
The parameters of NGC 5195 are listed in table 1.

\section{Observations and Data Reductions}

\subsection{NRO 45 m Observations}

We performed simultaneous HCN(1$-$0) and CO(1$-$0) 
observations toward the center of
NGC 5195 using the NRO 45 m telescope on 1996 December.
In order to avoid the degradation of
any beam pattern, beam efficiency, or pointing accuracy of the telescope,
the observed data were discarded 
if the wind velocity exceeded about 4 m s$^{\rm -1}$.
The full widths of the half-power beam (FWHP) were
$15''$ and $19''$ at the observed frequencies of CO and HCN, respectively.
The main-beam efficiency, $\eta_{\rm MB}$, was about 0.5 at the 3 mm band.
We used two cooled SIS mixer receivers, S80 and S100, 
equipped with side-band rejection filters,
and observed both the CO and HCN lines simultaneously.
The beam squint of the two receivers had been aligned to less than 2$''$.
Absolute pointing of the antenna was checked every hour
using SiO maser sources, and the pointing accuracy was better
than $\pm4''$ (peak-to-peak).
The typical system noise temperature for CO and HCN observations were
about 800 K and 300 K in the single side-band, respectively.
The sky emission was subtracted by position switching
with an off-source position at an offset in the azimuth of $8'$ 
from the observed position.
The spectra of CO and HCN emission were obtained with 2048 channels
acousto-optical spectrometers
of 250 MHz bandwidth, corresponding to a velocity coverage of
650 km s$^{\rm -1}$ for CO observations,
and 850 km s$^{\rm -1}$ for HCN observations.
After removing linear baselines from the spectra, adjacent channels
were averaged. The resultant velocity resolution is 10 km s$^{-1}$
for CO and 30 km s$^{-1}$ for HCN.
The 45 m telescope observations are summarized in table 2.

\subsection{NMA Observations}

NMA observations of CO(1$-$0) emission
in the central region of NGC 5195
were made during the period from 
1999 November to 2000 March. 
The NMA consists of six 10 m antennas equipped with cryogenically
cooled SIS mixer receivers. 
Three available array configurations (AB, C, and D) were used.
Due to the limitation of the minimum projected baseline length (10 m),
extended structures larger than about $50''$ in each channel map
were not sampled in the observations.
The receiver noise
temperatures at the 3 mm band were $20 - 40$ K in the double side-band, 
and the system noise temperatures in the single side-band
were about 800 K toward the zenith. 
The Ultra Wide-Band Correlator (UWBC; Okumura et al.\ 2000) was configured
to cover 512 MHz with 256 channels per baseline.
Side-band separation was achieved by 90$^\circ$ phase switching.
We observed the quasar 1418+546 every $\sim$ 30 min 
in order to calibrate the temporal variations of
the visibility amplitude and phase.
The passband across 256 channels was calibrated through observations
of a strong continuum source, 3C 279.
The flux density of the reference calibrator was determined
from comparisons with planets of known brightness temperatures.
It ranged from 0.5 to 0.7 Jy during the observation period,
and the uncertainty in the absolute flux scale was about $\pm$ 20 \%.
The raw data were calibrated and edited using the package UVPROC-II 
developed at NRO (Tsutsumi et al.\ 1997),
and Fourier-transformed using the AIPS task IMAGR.
We produced two CO data cubes with different imaging parameters;
one was maps using a natural weighting by applying a 80 k$\lambda$
taper to the visibilities, and longer baseline visibilities
($> 80$ k$\lambda$) were cut. This resulted in a moderate spatial
resolution of \timeform{4".7} $\times$ \timeform{3".2} 
(P.A.\ = $-$ $54^\circ$).
The other cube was made using a
robust weighting (robust parameter = 0) to achieve a high angular
resolution (\timeform{1".9} $\times$ \timeform{1".8}).
Hereafter, we refer to these two cubes as ``low resolution'' 
and ``high resolution'', respectively.
A conventional CLEAN method was applied 
to deconvolve the synthesized beam pattern.
Summaries of the observations and the parameters of the CO data
are listed in table 3.

\section{Results}

\subsection{NRO 45 m Observations}

We detected strong CO(1$-$0) emission in the center of NGC 5195,
whereas HCN(1$-$0) emission was found to be very weak.
The CO and HCN spectra are shown in figure 1,
and the quantities derived from these spectra are summarized
in table 2. 

The CO profile shows two velocity features, i.e., 
a $400 - 450$ km s$^{-1}$ component and
a $450 - 750$ km s$^{-1}$ component, which are mostly consistent
with the earlier CO detection (Sage 1990).
It has been suggested that the lower velocity component
is from the spiral arm of M 51, a companion of NGC 5195,
and that the higher velocity component is associated with
NGC 5195 itself. 
A high-resolution NMA map of CO(1$-$0)
emission, presented in the following section, confirms
this picture. 
The velocity-integrated CO intensity, $I$(CO) = 83 K km s$^{-1}$,
corresponds to a face-on molecular gas surface density
of about 430 $M_\odot$ pc$^{-2}$, adopting the formula
\begin{equation}
\Sigma_{\rm H_2} = 4.81 \times {\rm cos}\ i \cdot
\left( \frac{I_{\rm CO}}{\mbox{K km s$^{-1}$}} \right)
\left[ \frac{X_{\rm CO}}{3.0 \times 10^{20}\ \mbox{cm$^{-2}$ (K km s$^{-1}$)$^{-1}$} } \right] 
\end{equation}
and 
\begin{equation}
\Sigma_{\rm gas} = 1.36 \times \Sigma_{\rm H_2},
\end{equation}
where $i$ is the inclination of the disk ($i = 37^\circ$ here),
$I_{\rm CO}$ is the velocity-integrated CO intensity, and
$X_{\rm CO}$ is the $N_{\rm H_2}/I_{\rm CO}$ conversion factor
(Scoville et al.\ 1987; Solomon et al.\ 1987).

We find that the HCN emission
in the center of NGC 5195 is very weak. 
Because the detection was 
about the 3 $\sigma$ level, it should be regarded
as tentative. 
The HCN-to-CO integrated intensity ratio
in brightness temperature scale, $R_{\rm HCN/CO}$, 
is given as about 0.02. 
Note that the beam sizes of CO and HCN observations
are slightly different ($15''$ and $19''$, respectively),
and one should note that such a difference could cause 
an error in the estimation of the intrinsic gas properties.

\subsection{NMA Observations}

\subsubsection{Molecular-gas distribution and kinematics in NGC 5195}

Figure 2 shows velocity channel maps of the low-resolution CO data cube 
in the central $50'' \times 50''$ (corresponding to 2.3 kpc at $D$ = 9.3 Mpc)
area of NGC 5195. CO emission is clearly detected ($> 4 \sigma$) 
over the velocity range from 487 to 737 km s$^{-1}$,
and presumably detected even in the 471 and 752 km s$^{-1}$ channels.

A velocity-integrated CO map (the high resolution data)
over the $471 - 752$ km s$^{-1}$ range
is shown in figure 3c, along with an intensity-weighted 
mean-velocity map in figure 3d. 
CO maps made from the low-resolution data cube are shown in figure 4.
These 0th and 1st moment maps were produced
using the AIPS task MOMNT, with a clip level of 1.5 $\sigma$
in each channel. 
We can immediately see a strong concentration of CO emission
toward the nucleus of NGC 5195. 
The azimuthally averaged radial distribution
of molecular gas shown in figure 5, which was calculated using the AIPS task
IRING, suggests that the peak gas mass per unit area 
including He and heavier elements, $\Sigma_{\rm gas}$,
reaches about $3.7\times10^3\ M_\odot$ pc$^{-2}$, 
adopting the equations (1) and (2). 
The $\Sigma_{\rm gas}$ derived here is much higher than
that from the 45 m data. This is solely due to the dilution effect;
the beam filling factor of molecular emission is very small.

The CO peak corresponds to the radio peak position,
and is mostly associated with the H$\alpha$ {\it absorption} region,
which can be seen in figure 3b. In this region, a hot dusty ring
has also been discovered by a B-15 $\mu$m color map (Block et al.\ 1997),
and which could also be a reminiscent of a past starburst event. 
The relationship between molecular gas and the past starburst
in the center of NGC 5195 is discussed in section 4.

The CO flux, $S_{\rm CO}$, within the central $r<500$ pc ($11''$) region
in figure 4 is 280 Jy km s$^{-1}$, and the total flux 
detected in the F.O.V. is about 380 Jy km s$^{-1}$.
These correspond to a molecular gas mass of $4.0\times10^8\ M_\odot$
and $5.4\times10^8\ M_\odot$, respectively, following the equations
\begin{equation}
M(\mbox{H$_2$}) = 1.2\times10^4 
\left( \frac{S_{\rm CO}}{\mbox{Jy km s$^{-1}$}} \right) 
\left( \frac{D}{\mbox{Mpc}} \right)^2
\left[ \frac{X_{\rm CO}}{3.0\times10^{20}\ \mbox{cm$^{-2}$ (K km s$^{-1}$)$^{-1}$}} \right]
\end{equation}
and
\begin{equation}
M(\mbox{gas}) = 1.36 \times M(\mbox{H$_2$}).
\end{equation}
The averaged gas surface density within the $r<500$ pc region is 
$\Sigma_{\rm gas}(r<500\ \mbox{pc}) = 5.0\times10^2\ M_\odot$ pc$^{-2}$.
This $\Sigma_{\rm gas}$ is comparable to those in a normal spiral survey
(Sakamoto et al.\ 1999b), which consists of 20 CO-luminous spiral galaxies 
with morphology types from Sa to Sd. 
Note that the number of S0 galaxies where high-resolution CO images
are available is still limited to date
(e.g., NGC 4701, Wrobel, Kenney 1992; NGC 404, Cepa et al.\ 1998; 
NGC 3593, Sakamoto et al.\ 1999a; NGC 7465, Kohno et al.\ 2001),
though a large number of CO detections have been reported
(e.g., Sage, Wrobel 1989; Thronson et al.\ 1989; Tacconi et al.\ 1991;
Taniguchi et al.\ 1994). Our results suggest that
some S0 galaxies can contain a large amount of molecular gas 
comparable to disk galaxies.

The CO velocity field in the low-resolution map (figure 4) 
seems to be dominated by circular rotation.
A simple fitting of the observed velocity field with the AIPS
task GAL gives a systemic velocity of about $628 \pm 5$ km s$^{-1}$
and a P.A.\ of the kinematic major axis of 
about $+106^\circ \pm 3^\circ$ (measured counterclockwise from north), 
although it can depend on the area where the fitting is made.
The P.A. is mostly consistent with the value in RC3 (table 1). 
On the other hand,
the systemic velocity derived here differs significantly from the
H\emissiontype{I} velocity listed in RC3, $465\pm10$ km s$^{-1}$.
Our CO measurements must be more accurate,
because high-resolution observations of 
H\emissiontype{I} do not show any significant
H\emissiontype{I} emission in the center of NGC 5195 (Rots et al.\ 1990), 
and the listed H\emissiontype{I} velocity in RC3 seems
to be contaminated by the H\emissiontype{I} from M 51.

A position-to-velocity map ($pv$ map) 
along the kinematic major axis (P.A. = +106$^\circ$) is
displayed in figure 6.
We find a steep rise of rotation velocity on the $pv$ map; 
it reaches about 160 km s$^{-1}$ at $r \sim 1''$ or 45 pc
on the plane of the galaxy, adopting $i = 37^{\circ}$.
Note that we see a velocity component near $V_{\rm LSR} \sim 430$ km s$^{-1}$,
which may be CO emission associated with M 51 (see 3.2.2).

The observed gas morphology is reminiscent of those in barred spiral galaxies
(e.g., Kenney et al.\ 1992; Sakamoto et al.\ 1995; Reynaud, Downes 1997;
Regan et al.\ 1999);
two offset ridges emanate from the central gas concentration,
and are connected to curved structures. 
A curved or spiral-like feature, which can often be a signature 
of the gas orbit resonance in barred galaxies, 
is particularly evident in the NW part.  
In fact, near-infrared (NIR) photometry clearly shows 
the presence of a large-scale stellar bar along the N--S direction 
(Smith et al.\ 1990; Thronson et al.\ 1991; 
Spillar et al.\ 1992; Block et al.\ 1994);
it is very natural to have an idea that they govern the gas distribution.
If we assume that these are bar-induced structures, indeed, 
it is possible to specify the direction 
of rotation on the plane of the galaxy. 
In figure 7, we suggest a possible configuration.
As can be seen in the figure, 
the northern part of the galaxy is suggested to be the near side, 
which is consistent with results 
obtained by assuming the outer spiral arms
seen in NIR are trailing (Block et al.\ 1994). 
In order to fully understand the gas distribution and kinematics
in the center of NGC 5195, detailed analyses on 
gas kinematics, including a comparison with the hot dusty ring
(Block et al.\ 1997), will be needed.

\subsubsection{Low-velocity component, possible CO emission from M 51}

It should be noted that low-level ($3-4$ $\sigma$) emission
can be seen near to the edge of the NMA F.O.V.\ (mostly the S--E portion
of the F.O.V.) in the channel maps of the low-resolution CO cube (figure 2) 
for the velocity range $V_{\rm LSR} = 378 - 440$ km s$^{-1}$. 
If we correct the primary beam attenuation near the edge of F.O.V.,
these low-velocity emissions could be about 200 mJy beam$^{-1}$
or 1 K at the observing beam 
($4.\hspace{-2pt}''7 \times 3.\hspace{-2pt}''2$).
The velocity range of these low-level CO emissions
is mostly the same as that of the low-velocity component detected at 
45 m (and NRAO 12 m by Sage 1990). We show the integrated intensity
map for the 378 to 440 km s$^{-1}$ range in figure 8. 
A peak near the S--E edge of the F.O.V.\ can be clearly seen in the figure,
and this could be CO emission from M 51. 
This low-level emission is clearly depicted by recent multi-pointing
imaging of the NGC 5195 region using the OVRO millimeter array 
(Aalto, Rydbeck 2001). 
Nevertheless, the flux ratio between 45 m and NMA data is very small
for this velocity range, as can be seen in figure 9, which
compares the 45 m and NMA spectra, although most
of the CO emission from NGC 5195 itself seems to be recovered
with the NMA observations. We therefore suggest that
the CO emission from M 51 seen as the low-velocity component 
would be spread over a wide area of arcminutes scale,
and that most of the low-velocity CO emission is resolved out 
in our interferometric observations.
This seems to be consistent with the $pv$ map
by Sage (1990), which shows extended emission over a few arcminutes
for the velocity component near $V_{\rm LSR} \sim 400$ km s$^{-1}$.

\section{Discussions}

The NMA observations of CO emission reveal that the post-starburst galaxy
NGC 5195 does contain a significant amount of molecular gas,
despite the fact that the current massive star formation is very inactive there;
the molecular gas mass per unit area reaches about
$\Sigma_{\rm gas} \sim 3.7 \times 10^3\ M_\odot$ pc$^{-2}$,
which is comparable to those in nearby starburst galaxies.
On the other hand, the star formation rate (SFR) per unit area, 
$\Sigma_{\rm SFR}$,
which was calculated from $2'' \times 4''$ aperture H$\alpha$ data
(Ho et al.\ 1997a), is only about 
$\Sigma_{\rm SFR} \sim 3.7 \times 10^{-8}\ M_\odot$ yr$^{-1}$ pc$^{-2}$
(table 1).
This SFR is smaller than those in typical starburst galaxies by an order of magnitude
(e.g., Kennicutt 1998a).
The resultant gas consumption time scale,
\begin{equation}
\tau_{\rm gas} = \frac{\Sigma_{\rm gas}}{\Sigma_{\rm SFR}},
\end{equation}
is about $\sim 9 \times 10^{10}$ yr, 
which is indeed comparable to $\tau_{\rm gas}$
in quiescent or normal galaxies,
and significantly longer than those in infrared-luminous starburst galaxies 
(e.g., Kennicutt 1998a).
In the following section we discuss the possible relationship
between the physical properties of molecular gas and star formation
in the center of NGC 5195.

\subsection{Decrease of Dense Molecular Gas in the Nucleus of the Post-Starburst Galaxy NGC 5195}

We observed a very low $R_{\rm HCN/CO}$ value of 0.02 in the center of NGC 5195.
This is smaller than those in starburst galaxies by a factor of $5 - 15$.
For instance, $R_{\rm HCN/CO}$ in NGC 253 is reported to be in the range
of $0.2 - 0.3$ (Helfer, Blitz 1993; Paglione et al.\ 1995; Sorai et al. 2000),
and starburst galaxies such as NGC 6946 and IC 342, which are less active
compared with NGC 253 in terms of FIR luminosities, 
show $R_{\rm HCN/CO}$ values
of about 0.1 (Downes et al. 1992; Helfer, Blitz 1997).
$R_{\rm HCN/CO}$ values in the circumnuclear starburst regions 
of NGC 1068 (Helfer, Blitz 1995) and NGC 6951 (Kohno et al.\ 1999)
are also about 0.1 or so.
 
Because the critical gas density for collisionally excitation of HCN(1$-$0)
emission ($n_{\rm H_2} > 10^4$ cm$^{-3}$) is much higher than that of
CO(1$-$0) ($n_{\rm H_2} \sim $ a few $\times$ $10^2$ cm$^{-3}$),
a comparison of the CO and HCN intensities is a measure of 
the gas density if both of the CO and HCN emission 
originate from the same volume.
In observations of galaxies, the observing beams are often too large
to resolve the individual cloud structures, and 
$R_{\rm HCN/CO}$ could indicate the fraction 
of dense molecular gas to the total (including diffuse) molecular gas
within the observing beam (Kohno et al.\ 1999).

With this interpretation of $R_{\rm HCN/CO}$ in galaxies,
our data on NGC 5195 and its comparison with starburst galaxies
would indicate that {\it the mass fraction of dense molecular components
to the total molecular gas}, which is traced by the $R_{\rm HCN/CO}$ values, 
{\it in the center of NGC 5195 is significantly small compared with 
that of starburst galaxies}.
We suggest that this would be a reason 
why NGC 5195 is in a post-starburst phase
despite the fact that it contains a large amount of molecular gas there; 
most of the dense molecular cores,
which are the very sites of massive star formation, have been
consumed away by starburst events which began about 1 Gyr before 
(see section 1).
Also, a burst of massive star formation can no longer last,
although a large amount of {\it diffuse} molecular gas remains there.

We should note that the $R_{\rm HCN/CO}$ values often depend on
the aperture size of the observations, and attention must be paid
to derive the intrinsic properties of molecular gas 
from the $R_{\rm HCN/CO}$ values.
(e.g., Helfer, Blitz 1993; Jackson et al.\ 1996).
For instance, $R_{\rm HCN/CO}$ in the center of M 51 is 0.033 
(Helfer, Blitz 1993)
by the NRAO 12 m telescope ($55''$ for CO, and $71''$ for HCN), 
whereas it becomes about 0.08 (Sorai et al.\ 2002) 
by the NRO 45 m telescope ($15''$ and $19''$ for CO and HCN, respectively).
In higher resolution ($7''$) observations, it reaches about 0.3 or more
(Kohno et al.\ 1996). 
These effects are mostly due to the compactness
of the HCN emitting regions, and therefore high-resolution observations
are essential in order to address the relationship between dense molecular 
gas and massive star-forming regions.
In the case of our NGC 5195 observations, the observing beams 
($15''$ for CO and $19''$ for HCN) correspond to
aperture areas of $r<340$ pc and $r<430$ pc for CO and HCN, respectively.
Small $R_{\rm HCN/CO}$ values of less than 0.05 
have been reported in various galaxies
(e.g., Helfer, Blitz 1993; Aalto et al. 1995), 
but are mostly observed with a much larger beam size ($\sim 1'$) than ours, 
corresponding to a few $\times$ kpc aperture area in their samples.
We therefore suggest that the observed small $R_{\rm HCN/CO}$ value
in the central a few $\times$ 100 pc of NGC 5195 
is not the aperture size effect
but reflects the intrinsic property of molecular gas, i.e., a decrease
of dense molecular gas there.
Our interpretation of a low $R_{\rm HCN/CO}$ ratio seems to be
supported by a low CO(2$-$1)/CO(1$-$0) ratio in NGC 5195; 
Sage (1990) observed CO(2$-$1) and CO(1$-$0) lines 
toward the center of NGC 5195 using the NRAO 12 m telescope,
and found that the CO(2$-$1)/CO(1$-$0) ratio is as small as 0.5.
They concluded that the molecular clouds in NGC 5195 are {\it less dense}
than those in typical spiral galaxies.

\subsection{Gravitationally Stable Molecular Gas Disk in the Center of NGC 5195}

Given the close relationship between dense molecular gas 
and massive star formation/starburst, 
what kind of physical processes can play a major role
in the formation of {\it dense} molecular clouds from a {\it low-density}
molecular medium? One of the key processes could be gravitational instabilities
of molecular gas; it is well known that SFR in the disk regions
($\sim$ a few kpc scale) of galaxies
can be described by means of the Schmidt law, 
$\Sigma_{\rm SFR} \propto \Sigma_{\rm gas}^{1.4\pm0.15}$, where
$\Sigma_{\rm SFR}$ is the SFR per unit area, and $\Sigma_{\rm gas}$ is
the gas mass per unit area (Kennicutt 1998a).
Here it should be noted that the Schmidt law is valid 
only if the gas in the disk is gravitationally {\it unstable},
i.e., the gas surface density, $\Sigma_{\rm gas}$, exceeds the critical 
gas surface density, $\Sigma_{\rm crit}$, given by
\begin{equation}
\Sigma_{\rm crit}
= \alpha \frac{\sigma_v \kappa}{\pi G}
= 73.9 \times \left(\frac{\alpha}{1}\right) \left(\frac{\sigma_v}{\mbox{km s$^{-
1}$}}\right)
\left(\frac{\kappa}{\mbox{km s$^{-1}$ pc$^{-1}$}}\right)
M_\odot \mbox{\ pc$^{-2}$},
\end{equation}
where $\alpha$ is a dimensionless constant close to unity\footnote{
In the case of a purely gaseous disk, $\alpha$ is unity.
Yet $\alpha$ could be less than
unity for a realistic two-fluid stability condition
because the interaction 
between the stellar and gaseous disks could act 
to destabilize the gas disk (Jog, Solomon 1984).
Martin and Kennicutt (2001) has shown that 
the best-fit value of $\alpha$ is 0.69 $\pm$ 0.2;
we use this in the following calculation.},
$\sigma_v$ is the velocity dispersion in the radial direction, and
$\kappa$ is the epicyclic frequency,
\begin{equation}
\kappa = \left\{ 2 \frac{v(r)}{r} \left[\frac{v(r)}{r} + \frac{dv(r)}{dr}\right] \right\}^{0.5},
\end{equation}
where $r$ is the distance from the center and
$v(r)$ is the rotation velocity at the radius of $r$
after correcting the inclination as 1/sin $i$.
If the gas surface density, $\Sigma_{\rm gas}$,
exceeds the threshold, $\Sigma_{\rm crit}$, then
a uniform gas disk is unstable to form rings or clumps 
which can ultimately collapse into dense molecular gas fragments 
and form stars, eventually.
This criterion, for local gravitational stability
in thin isothermal disks, is often expressed as
$Q = \Sigma_{\rm crit}/\Sigma_{\rm gas}$ (Toomre's $Q$ parameter; 
Toomre 1964); $Q>1$ ($Q<1$) means the gas disk is stable (unstable).

Surprisingly, this criterion has been successfully applied
to much smaller scale disks, i.e., inner a few $\times$ 100 pc regions
of galaxies, to explain the star-formation properties of
starbursts (e.g., NGC 3504, Kenney et al.\ 1993) and rather normal
(e.g., NGC 4414, Sakamoto 1996) galaxies. It is therefore natural
to expect that the gravitational instabilities of molecular gas
would be related to dense molecular gas formation.
In fact, Kohno et al.\ (1999) reported that $R_{\rm HCN/CO}$
is enhanced in regions where the $Q$ parameter is almost equal to 
or smaller than unity, suggesting a connection between
dense gas formation and gravitational instabilities of molecular gas.
 
In order to assess the stability of a molecular gas disk
in the center of NGC 5195, we calculated $\Sigma_{\rm crit}$
in NGC 5195 using the above equations.
The circular rotation velocities, obtained from the $pv$ map in figure 6,
were employed to compute the epicyclic frequency, $\kappa$. 
Note that our interest is just at the center of the gas disk, and
that the gas motion can be treated as a rigid rotation in this region.
In this case, the epicyclic frequency can be written
just as
\begin{equation}
\kappa = \sqrt{2} \left[\frac{v(r)}{r}\right].
\end{equation}
From the $pv$ map in figure 6, we suppose that
the largest circular rotation velocity occurs at about
$r \sim 1''$ or 50 pc
and the $v(r)_{\rm max}$ is
about $100 \mbox{\ km s$^{-1}$}/{\rm sin}\ i = 160$ km s$^{-1}$,
yielding a $\kappa$ of 4.5 within $r<50$ pc. 
Note that $\kappa$ must be a lower limit because
the radius of $1''$ where the rotation rises to its peak
could be an upper limit, given our limited angular resolution
($\sim 2''$).
Concerning the velocity dispersion of molecular gas,
we here assumed an intrinsic gas velocity dispersion,
$\sigma_v$, of 30 km s$^{-1}$,
which is similar to those in NGC 3504 (Kenney et al.\ 1993). 
This is because a precise extraction of intrinsic gas velocity dispersion 
from the observed CO line width (figure 9) is not easy near the center
of galaxies due to contamination of
the steep velocity gradient within the observing beam.
Consequently, we find a critical density, 
$\Sigma_{\rm crit}$, of $6.9\times10^3\ M_\odot$
pc$^{-2}$, and that the $Q = \Sigma_{\rm crit}/\Sigma_{\rm gas}$
is larger than unity (1.9) in the center of NGC 5195.
Hence, the molecular gas disk in the center of NGC 5195 would be 
gravitationally {\it stable}; we suggest that this may be the
reason why dense molecular gas is not formed from the large amount
of diffuse molecular gas in NGC 5195, since they are {\it too stable
to form dense molecular clouds via gravitational instabilities
of a diffuse molecular medium}. 

Although there are some sources of error,
we regard this conclusion as begin rather robust.
Though one of the major uncertainties of the $Q$ parameter comes from
$X_{\rm CO}$, recent studies
tend to show {\it smaller} values than the $X_{\rm CO}$ we adopted
here by a factor of $2-3$ or more 
in the central regions of various galaxies 
(e.g., Nakai, Kuno 1995; \cite{reg00}),
including the Galactic Center (e.g., Oka et al.\ 1998; Dahmen et al.\ 1998).
Even in the solar neighborhood, the latest extensive CO survey data 
of the Milky Way suggest a $X_{\rm CO}$ of
$1.8 \times 10^{20}$ cm$^{-2}$ (K km s$^{-1}$)$^{-1}$
(Dame et al.\ 2001), which is indeed smaller than the $X_{\rm CO}$ 
adopted in our calculation.
Thus the uncertainty of $X_{\rm CO}$ does not weaken our conclusion.
Although another source of error is the velocity dispersion, 
a large velocity dispersion of up to a few $\times$ 10 km s$^{-1}$ 
can be observed even in rather quiescent galaxies; 
the Galactic Center is the case (e.g., Spergel, Blitz 1992).
Given the finite spatial resolution, the epicycle frequency,
$\kappa$, could be a lower limit, 
indicating that the critical density can be even larger.
It increases the $Q$ value then.
All of these error sources thus indicate that the molecular gas
in the center of NGC 5195 is indeed supercritical ($Q>1$)
despite the large molecular gas mass there.

Supercritical ($Q>1$) molecular gas
in post-starburst galaxies has also been reported at the centers
of NGC 4736 (Shioya et al.\ 1996) and NGC 7331 (Tosaki, Shioya 1997);
both of them are prototypical post-starbursts 
(see Taniguchi et al.\ 1996 and reference therein for NGC 4736,
and see Ohyama, Taniguchi 1996 for NGC 7331). Gravitational instabilities
of molecular gas may, therefore, be closely related to 
dense molecular gas formation and, in turn, massive star-formation 
in galaxies.

\subsection{Star-Formation Properties in Early-Type Galaxies}

It should also be addressed why the molecular gas disk in NGC 5195
is supercritical. As mentioned above, the gas surface density,
$\Sigma_{\rm gas}$, is rather comparable to those in nuclear starburst
regions of galaxies. The major difference between starbursts and
the post-starburst galaxy NGC 5195 may be the very high
critical density for the gravitational instability;
the deduced $\Sigma_{\rm crit}$ of $6.9\times10^3\ M_\odot$ pc$^{-2}$
is significantly high compared with the central gas surface density
in disk galaxies (e.g., Sakamoto et al.\ 1999b).

This very high threshold may be caused by the large mass concentration
in early-type galaxies (Kennicutt 1989).
The large concentration of matter
results in a steep rise of the rotation velocity, as observed in NGC 5195. 
This increases the threshold density, because
$\kappa$ becomes very high in this case.
In other words, molecular gas at the centers of early-type galaxies
tends to be stabilized due to a small gas mass-to-dynamical mass ratio,
$M_{\rm gas}/M_{\rm dyn}$, 
even if a significant molecular gas mass concentration,
comparable to late-type disk galaxies, occurs (e.g., Young, Scoville 1991).

These properties observed in the S0 galaxy NGC 5195 
may explain the difference in the star-formation properties
along the Hubble sequence.
As reviewed by Kennicutt (1998b),
it is now established that 
the detection frequency of nuclear star formation is lower
in early-type galaxies than in late-type ones
(e.g., the detection rate is about 8\% in S0, whereas it increases
about 50\% in Sb and reaches 80\% in Sc-Im; Ho et al.\ 1997b),
whereas the strength of nuclear star formation shows
the opposite trend (i.e., the average extinction-corrected
H$\alpha$ luminosities in S0--Sbc galaxies is
nine-times higher than in Sc galaxies; Ho et al.\ 1997b).
The high critical density due to the high mass concentration
in early-type galaxies may explain these apparently
incoherent trends; in early-type galaxies,
the occurrence of nuclear star formation is low because
nuclear star formation cannot occur until
the gas surface density exceeds the higher threshold density
compared with late-type galaxies. 
In order to exceed the threshold, it may take a longer time
or many triggers to transport the larger amount of
molecular gas into the centers of early-type galaxies.
Once the gas surface density exceeds the threshold,
however, the star-formation rate must be very high,
because the gas surface density already reaches a very high 
value compared with late-type galaxies (Kennicutt 1989).
Our high-resolution CO observations of the early type
galaxy NGC 5195 do support this scenario.
A CO survey of early-type galaxies
using millimeter-wave interferometers
and a comparison of the gas surface and the threshold
densities with those in late-type spirals 
must be valuable to statistically test the picture
described above 
regarding the difference in star formation along the Hubble sequence.

\section{Summary}

We present high-resolution CO(1$-$0) and HCN(1$-$0) observations
of the Post-Starburst galaxy NGC 5195 made with the
Nobeyama Millimeter Array and the NRO 45 m telescope.
Our results and conclusions are summarized as follows:

\begin{enumerate}

\item{We detected strong CO emission in the center of NGC 5195, 
whereas HCN emission was found to be very weak. The HCN-to-CO 
integrated intensity ratio in brightness temperature scale,
$R_{\rm HCN/CO}$, is about 0.02. This $R_{\rm HCN/CO}$ is
very small compared with starburst galaxies by a factor of about $5-15$.
}

\item{Our NMA map of CO emission reveals a concentrated CO distribution
toward the nucleus, where little massive star formation is observed;
The molecular gas mass per unit area reaches about
$\Sigma_{\rm gas} \sim 3.7 \times 10^3\ M_\odot$ pc$^{-2}$,
which is comparable to those in nearby starburst galaxies,
whereas the massive star formation rate in the center
is smaller than those in typical starburst galaxies by orders of magnitude.
The resultant gas consumption time scale, $\tau_{\rm gas}$, is 
about $\sim 9 \times 10^{10}$ yr, 
showing very small star-formation efficiency
in the center of NGC 5195 compared with that of starburst galaxies.}

\item{We suggest that the very small $R_{\rm HCN/CO}$ value 
could explain the reason why the starburst in NGC 5195 has ceased;
most of the dense molecular cores,
which are the very sites of massive star-formation, have been
consumed away by starburst events which began about 1 Gyr before.
A burst of massive star formation can no longer last,
although a large amount of {\it diffuse} molecular gas remains there.
}

\item{We have found a steep rise of rotation velocity 
toward the center of NGC 5195.
It reaches about 160 km s$^{-1}$ at $r \sim 50$ pc
on the plane of the galaxy.
As a consequence, the critical gas surface density for gravitational
instability of the gas disk is very high ($\Sigma_{\rm crit} \sim 6.9 \times
10^3$ $M_\odot$ pc$^{-2}$), and we suggest that the gas disk in the center
of NGC 5195 is gravitationally stable.
Although NGC 5195 contains
large quantities of {\it low density} molecular gas in the center,
they must be too stable to form dense molecular clouds via
gravitational instabilities of molecular gas.}

\item{The deduced very high threshold density seems to be due to 
a high mass concentration in the center of NGC 5195, 
and may be characteristic of early-type galaxies. 
If this is the case, the known trends on the 
occurrence and luminosity of nuclear star formation 
along the Hubble sequence can be explained naturally.}

\end{enumerate}

\bigskip

We would like to thank the referee for the helpful comments 
that have improved this paper.
We acknowledge the NRO staff for operating the telescopes
and continuous efforts to improve the performance of the instruments.

\clearpage

%%% Fig.1
\begin{figure}
\begin{center}
\FigureFile(85mm,80mm){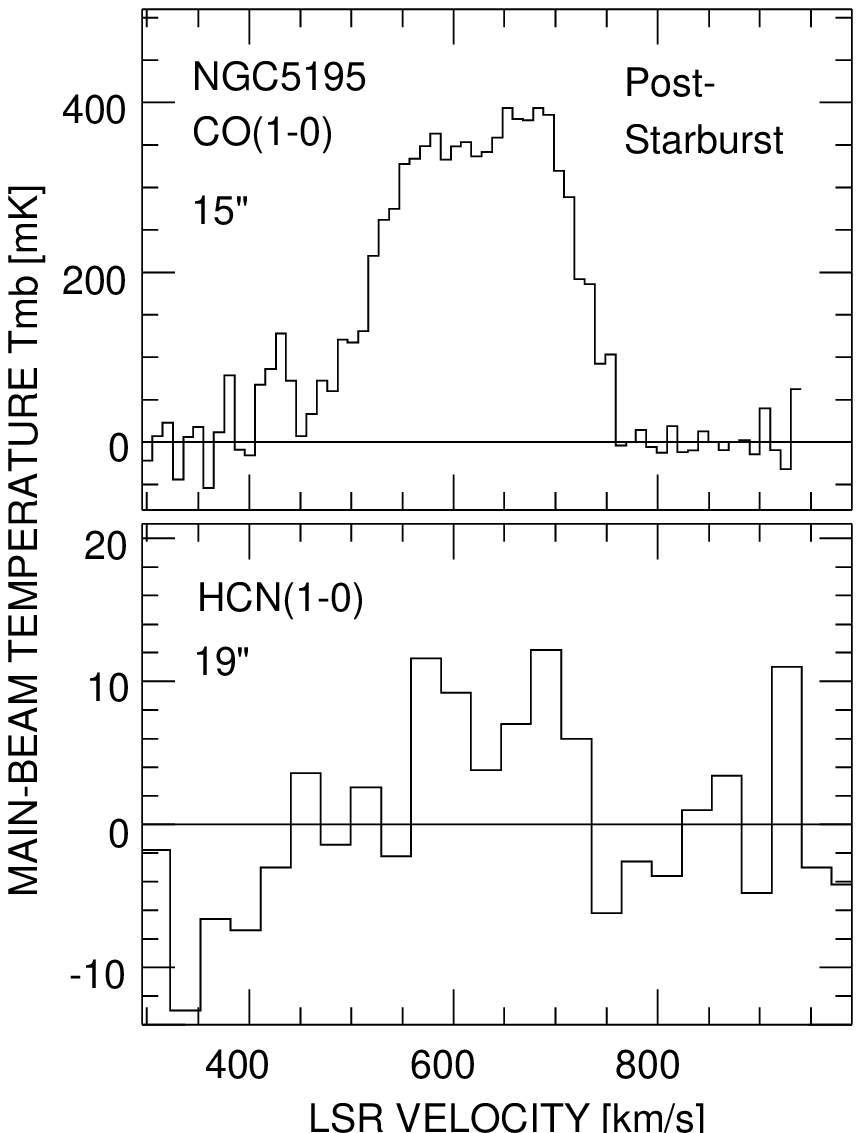}
\end{center}
\caption{CO(1$-$0) and HCN(1$-$0) spectra in the center of NGC 5195
obtained with the NRO 45 m telescope. The beam sizes (HPBW)
of the CO and HCN data are $15''$ and $19''$, respectively.
}\label{fig:45spectra}
\end{figure}

%%% Fig.2
\begin{figure}
\begin{center}
\FigureFile(160,160){fig2a.eps}
\end{center}
\caption{CO(1$-$0) emission in 15.6 km s$^{-1}$ wide channels
from the central $50'' \times 50''$ region
(2.25 kpc at $D$ = 9.3 Mpc) of NGC 5195 observed with the NMA.
The synthesized beam is \timeform{4".7} $\times$ \timeform{3".2}
($210 \times 140$ pc) with a P.A. of $-54^{\circ}$ (the low resolution map).
The contour levels are $-4$, $-2$, 2, 4, 6, $\cdots$, 16, and 18 $\sigma$,
where 1 $\sigma$ = 23 mJy beam$^{-1}$ or 140 mK in $T_{\rm b}$.
Negatives are dashed, and zero contour is omitted. 
The cross indicates the position of the phase center, which corresponds
to the peak position of 6 cm radio continuum. NMA field of view
($60''$) is also indicated.
The attenuation of the 10 m dish is not corrected in these maps.
}\label{fig:chanmap1}
\end{figure}

%%% Fig.2 (cont.)
\addtocounter{figure}{-1}
\begin{figure*}
 \begin{center}
  \FigureFile(160mm,200mm){fig2b.eps}
 \end{center}
\caption{(Continued)}
\label{fig:chanmap2}
\end{figure*}

%%% Fig.3
\begin{figure}
\begin{center}
\FigureFile(160,160){fig3.eps}
\end{center}
\caption{High-resolution CO(1$-$0) integrated intensity 
and mean-velocity maps
along with optical (broad band and narrow band) images of NGC 5195;
(a) DSS $B$-band image of NGC 5195. The central $3' \times 2.\hspace{-2pt}'5$
is shown, and the NMA field of view ($60''$) is superposed on the picture
as a dashed circle. The solid box corresponds to the CO map region below.
(b) Pseudo-color H$\alpha$ + [N\emissiontype{II}] image of NGC 5195 
(Greenawalt et al.\ 1998).  
The contour levels are $-300$, $-100$, $-30$, $-10$, 
10, 30, 100, and 300 $\sigma$,
where 
1 $\sigma$ = $2.4 \times 10^{17}$ erg s$^{-1}$ cm$^{-2}$ arcsec$^{-2}$.
Negatives are dashed. NMA field of view is indicated (dashed circle).
Note the strong absorption toward the center of NGC 5195 (the central
cross) with a diameter of about $10''$ or 450 pc.
(c) High-resolution integrated intensity map of CO(1$-$0) emission 
from the central \timeform{30"} $\times$ \timeform{25"} region
(1.35 kpc $\times$ 1.13 kpc) of NGC 5195.
The synthesized beam is \timeform{1".9} $\times$ \timeform{1".8}
(86 pc $\times$ 81 pc) with a P.A. of $-48^{\circ}$.
The contour levels are 2, 4, 6, 8, 10, 15, and 20 $\sigma$,
where 1 $\sigma$ = 1.5 Jy beam$^{-1}$ km s$^{-1}$
or 39.4 K km s$^{-1}$ in $T_{\rm b}$.
The 1 $\sigma$ corresponds to the face-on
gas surface density, $\Sigma_{\rm gas}$, of 190 $M_\odot$ pc$^{-2}$.
(d) Intensity-weighted mean velocity map of CO(1$-$0) emission
The contour interval is 20 km s$^{-1}$.
}\label{fig:colormaps}
\end{figure}

%%% Fig.4
\begin{figure}
\begin{center}
\FigureFile(160,120){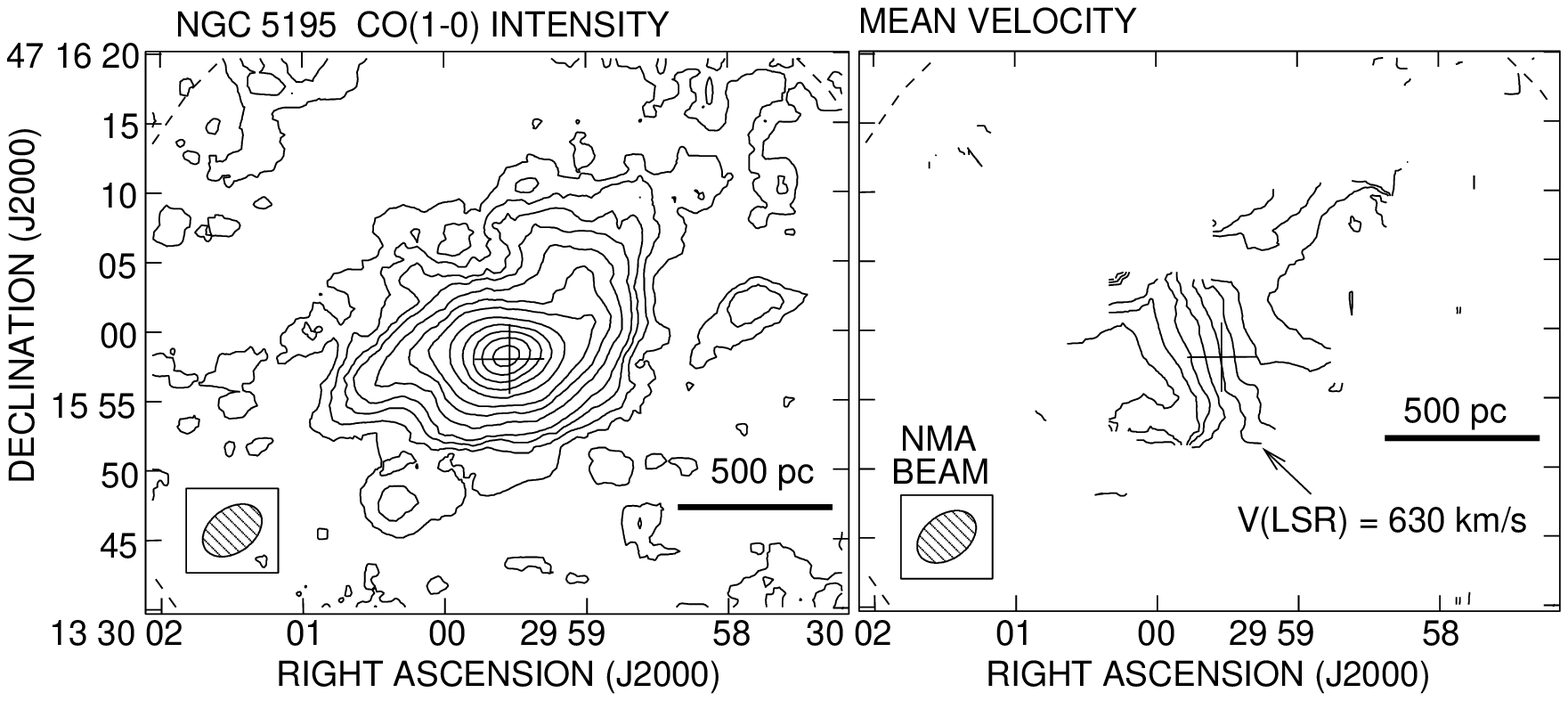}
\end{center}
\caption{
Low resolution CO data of NGC 5195.
(left) Integrated intensity map of CO.
The synthesized beam is 
$4.\hspace{-2pt}''7 \times 3.\hspace{-2pt}''2$
(210 pc $\times$ 140 pc) with a P.A. of $-54^{\circ}$.
The contour levels are 1.5, 3, 4.5, 6, 9, 12, 15, 20, 25, 30, 35, 
40, and 45 $\sigma$, where 1 $\sigma$ = 1.5 Jy beam$^{-1}$ km s$^{-1}$
or 9.3 K km s$^{-1}$ in $T_{\rm b}$. This corresponds to a face-on
gas surface density, $\Sigma_{\rm gas}$, of 48.8 $M_\odot$ pc$^{-2}$.
(right) Intensity weighted mean velocity map of CO.
The contour interval is 20 km s$^{-1}$.
}\label{fig:cohireso}
\end{figure}

%%% Fig.5
\begin{figure}
\begin{center}
\FigureFile(85,70){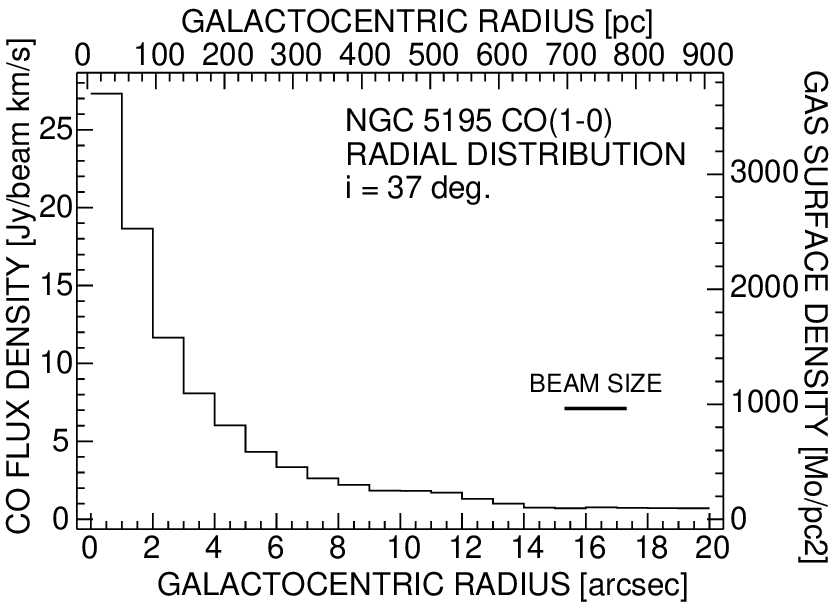}
\end{center}
\caption{
Radial distribution of CO emission in the central $r<25''$ (1.1 kpc) region
of NGC 5195. The high-resolution CO integrated intensity map (figure 4) 
is azimuthally averaged
over the successive annuli with a $1''$ = 45 pc width. Corrections
for the inclination and primary beam attenuation are applied. 
The face-on gas surface density was calculated 
by adopting a conversion factor of $X_{\rm CO} =
3.0 \times 10^{20}$ cm$^{-2}$ (K km s$^{-1}$)$^{-1}$, including
heavier elements.
A strong gas concentration toward the nucleus is evident.
}\label{fig:radialdistribution}
\end{figure}

%%% Fig.6
\begin{figure}
\begin{center}
\FigureFile(85,70){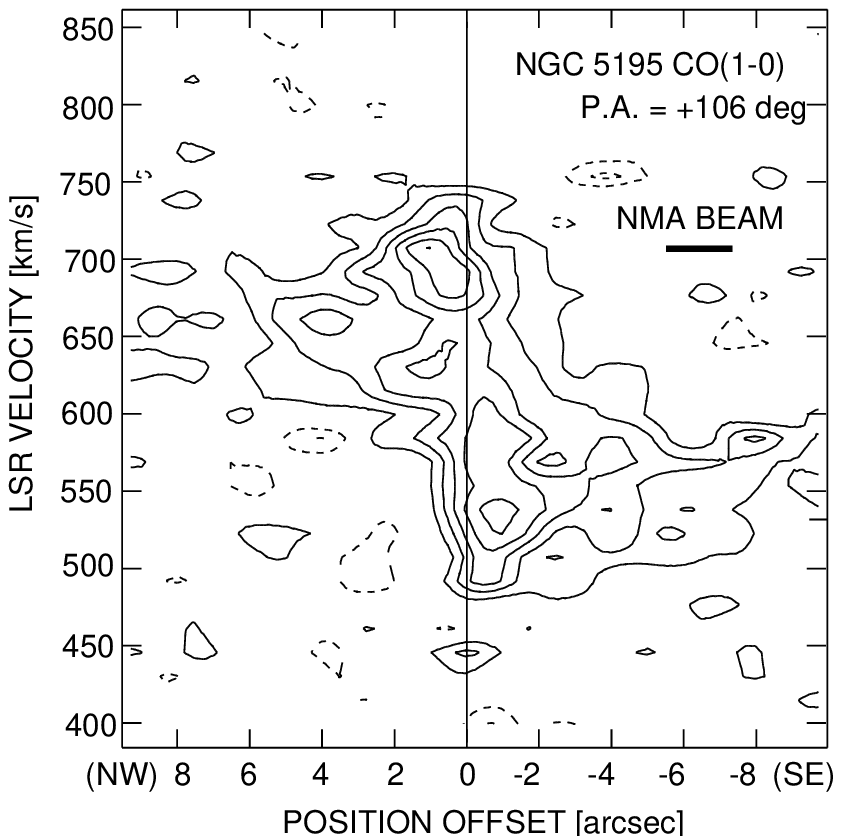}
\end{center}
\caption{
Position-to-Velocity ($pv$) map of CO emission along the P.A.
of +106$^\circ$ (measured counterclockwise from N), 
i.e., the kinematical major axis determined from
the CO velocity field. This $pv$ map is derived from the high-resolution
CO data cube. A correction of inclination was not applied,
whereas a primary beam correction was applied. The origin of position offset
corresponds to $\alpha$(J2000) = 
$13^{\rm h}29^{\rm m}59^{\rm s}\hspace{-5pt}.\hspace{2pt}54$ and 
$\delta$(J2000) = $+47^{\circ}15'58.\hspace{-2pt}''0$.
The contour levels are $-3$, $-1.5$, 1.5, 3, $\cdots$, 9 $\sigma$,
where 1 $\sigma$ = 22 mJy beam$^{-1}$.
}\label{fig:pv}
\end{figure}

%%% Fig. 7
\begin{figure}
\begin{center}
\FigureFile(85,70){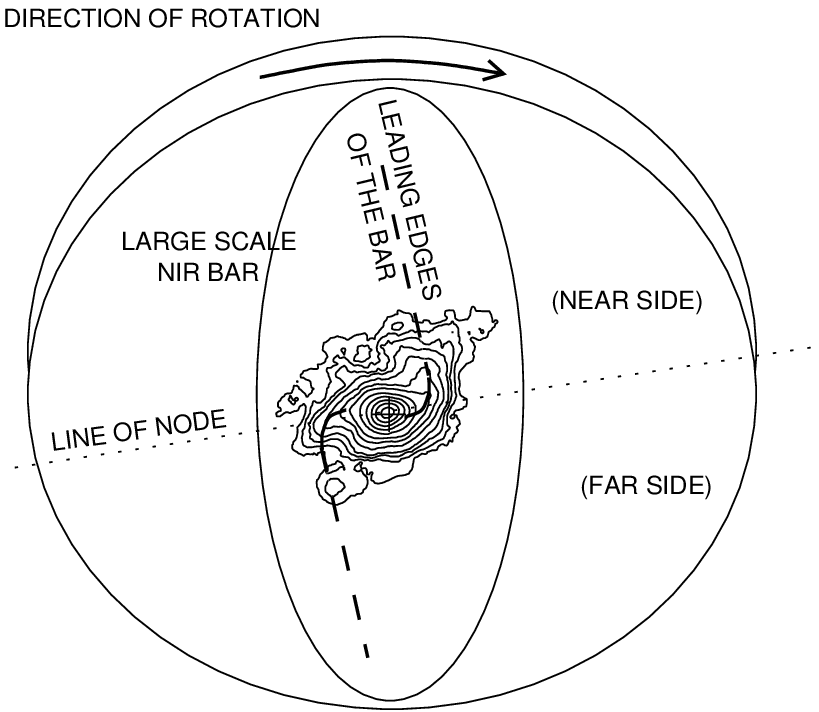}
\end{center}
\caption{
Schematic view of NGC 5195. This is based on the assumption that
the observed CO structures, such as two offset ridges emanating
from the nuclear peak and curved/spiral-like structures, 
are driven by a large-scale stellar bar,
which is clearly identified in the NIR bands.
}\label{fig:schematicview}
\end{figure}

%%% Fig. 8
\begin{figure}
\begin{center}
\FigureFile(85,70){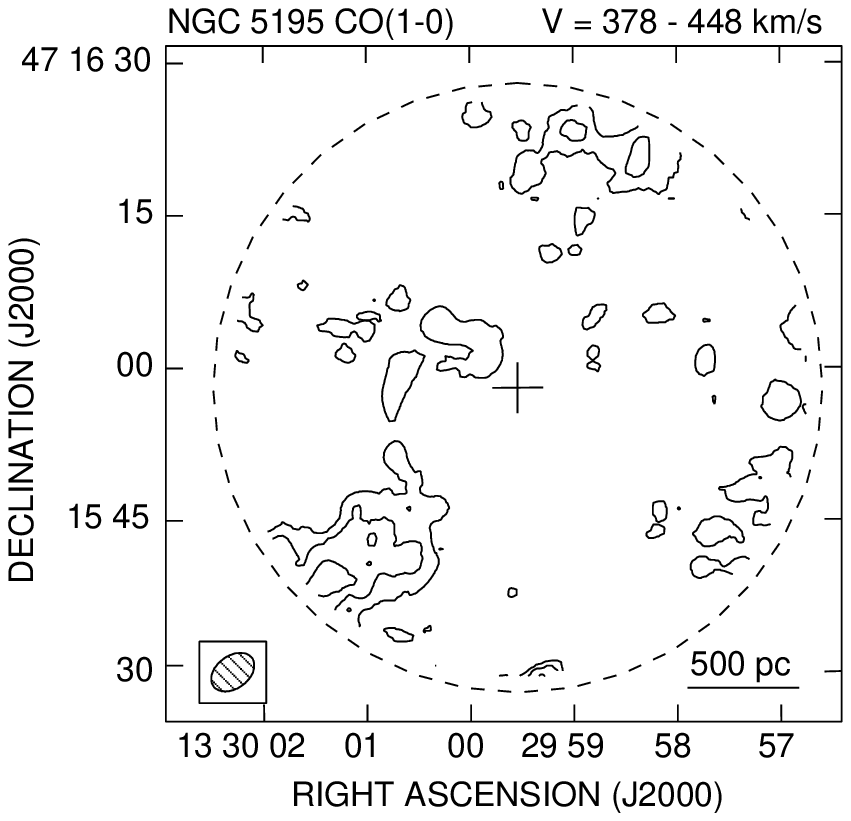}
\end{center}
\caption{
Integrated intensity map of CO emission over the velocity range
$V_{\rm LSR} = 378 - 448$ km s$^{-1}$, showing CO emission possibly from
a spiral arm of M 51. This map was produced from the low-resolution CO cube.
The contour levels are 2, 4, and 6 $\sigma$,
where 1 $\sigma$ = 0.80 Jy beam$^{-1}$ km s$^{-1}$ or 4.9 K km s$^{-1}$
in $T_{\rm b}$,
corresponding to a face-on gas surface density of
$\Sigma_{\rm gas} = 26\ M_\odot$ pc$^{-2}$,
including heavier elements as 
$\Sigma_{\rm gas} = 1.36 \times \Sigma_{\rm H_2}$.
The primary beam attenuation was corrected.
}\label{fig:lowvelocity}
\end{figure}

%%% Fig. 9
\begin{figure}
\begin{center}
\FigureFile(85,70){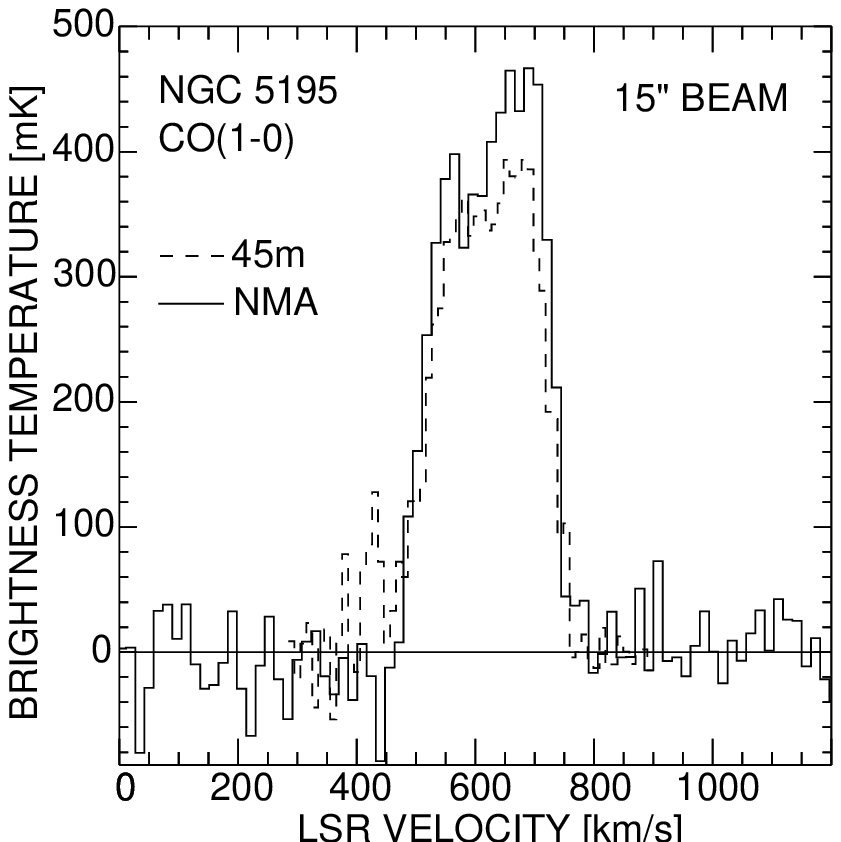}
\end{center}
\caption{
Comparison of the CO spectra taken with the NMA and 45 m. The NMA spectrum
was made after convolving the low-resolution CO data cube to the same beam size
as the 45 m beam ($15''$). Note that low-velocity component
(near 400 km s$^{-1}$) seen in the 45 m spectrum is not evident
in the NMA cube, suggesting that this velocity component distributes widely
beyond the NMA F.O.V. and would be mostly resolved out in the NMA data.
}\label{fig:nmavs45}
\end{figure}

%%% Fig. 10
\begin{figure}
\begin{center}
\FigureFile(85,60){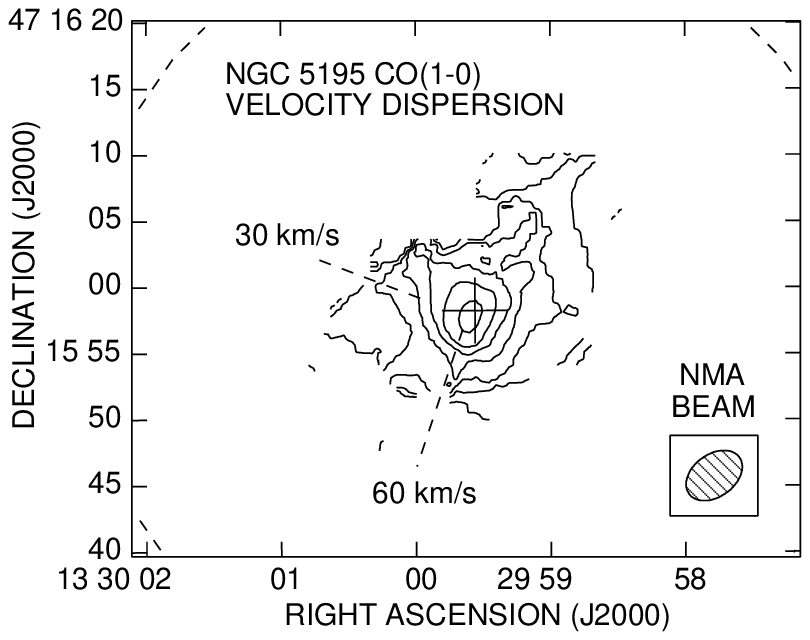}
\end{center}
\caption{
Intensity-weighted velocity dispersion map of the CO emission
in NGC 5195, which was computed by calculating the
second-order moment of intensity from the low-resolution CO cube.
The contour interval is 10 km s$^{-1}$.
Note that this map contains both the intrinsic gas velocity
dispersion and the gradient of rotation velocity
across the observing beam.
}\label{fig:mom2}
\end{figure}

%%%%%%%%%%%
% Table 1

\begin{table*}[hbt]
%\small
\begin{center}
Table~1.\hspace{4pt}Parameters of NGC 5195.\\
\end{center}
\vspace{6pt}
\begin{tabular*}{\textwidth}{@{\hspace{\tabcolsep}
\extracolsep{\fill}}p{12pc}ccc}
\hline\hline\\[-6pt]
Parameter & Value & Reference & \\
[4pt]\hline\\[-6pt]
Morphology \dotfill & IA0 pec  & (a) \\
                    & SB0 pec  & (b) \\
Size ($D_{\rm 25} \times d_{\rm 25}$) \dotfill &
$5.\hspace{-2pt}'8 \times 4.\hspace{-2pt}'6$  & (a) \\
Nuclear activity \dotfill & L2: & (c) \\
                          & post-starburst & (d), (e) \\
Position of nucleus \dotfill & & (f) \\
\ \ $\alpha$(J2000) & $13^{\rm h}29^{\rm m}59^{\rm s}\hspace{-5pt}.\hspace{2pt}5
4$  &  \\
\ \ $\alpha$(J2000) & $+47^{\circ}15'58.\hspace{-2pt}''0$ &  \\
Position angle \dotfill & $+101^\circ$ (counterclockwise from N) & (a) \\
Inclination angle \dotfill & $37^\circ$ (face-on = $0^\circ$) & (a) \\
Distance \dotfill & 9.3 Mpc & (g) \\
Liner scale \dotfill & 45.1 pc arcsec$^{-1}$ &  \\
H$\alpha$ luminosity ($2''\times 4''$ area) \dotfill & $8.7 \times 10^{37}$ erg
s$^{-1}$ & (c) \\
SFR per unit area$^{\dag}$ $\Sigma_{\rm SFR}$ \dotfill & $3.7 \times 10^{-8}$ $M
_\odot$ pc$^{-2}$ yr$^{-1}$ &
\\[4pt]
\hline
\end{tabular*}

\vspace{6pt}

\noindent
References: (a) de Vaucouleurs et al.\ 1991, RC3;
(b) Sandage, Tammann 1981, RSA;
(c) Ho et al.\ 1997a; (d) Rieke 1988; (e) Boulade et al.\ 1996;
(f) Hummel et al.\ 1987; (g) Tully 1988.

\vspace{6pt}

\noindent
$^{\dag}$ Star formation rate (SFR) was derived as
$L_{\rm H\alpha}/(1.12 \times 10^{41}\ \mbox{\ erg s$^{-1}$})$
in $M_\odot$ yr$^{-1}$ (Kennicutt 1983).
No correction for extinction was made.

\end{table*}

%%%%%%%%%%%
% Table 2

\begin{table*}
%\small
\begin{center}
Table~2.\hspace{4pt}NRO 45 m observations.\\
\end{center}
\vspace{6pt}
\begin{tabular*}{\textwidth}{@{\hspace{\tabcolsep}
\extracolsep{\fill}}p{20pc}ccc}
\hline\hline\\[-6pt]
Parameter & CO(1$-$0)  & HCN(1$-$0) & \\
[4pt]\hline\\[-6pt]
Beamsize (FWHP) \dotfill & $15''$ & $19''$ \\
Main beam efficiency $\eta_{\rm MB}$ \dotfill & 0.5 & 0.5 \\
Velocity resolution $\Delta v$ [km s$^{-1}$] \dotfill & 10 & 30 \\
Peak temperature in $T_{\rm MB}$ [mK]    \dotfill & $4.0\times10^2 \pm 24$ & $12
 \pm 5.2$ \\
Integrated intensity$^{\dag}$ $\int T_{\rm MB}(v) dv$ [K km s$^{-1}$] \dotfill &
$83 \pm 1.4$ & $1.5 \pm 0.60$ \\
Luminosity $L'$ [K km s$^{-1}$ pc$^2$] \dotfill & $3.8\times10^7$ & $1.3\times10
^6$
\\[4pt]
\hline
\end{tabular*}

\vspace{6pt}

\noindent
$^{\dag}$ Integrated over the velocity range of $V_{\rm LSR} = 450 - 750$ km s$^
{-1}$.
\end{table*}

%%%%%%%%%%%
% Table 3

\begin{table*}
%\small
\begin{center}
Table~3.\hspace{4pt}NMA observations.\\
\end{center}
\vspace{6pt}
\begin{tabular*}{\textwidth}{@{\hspace{\tabcolsep}
\extracolsep{\fill}}p{20pc}cc}
\hline\hline\\[-6pt]
Parameter & Value & \\
[4pt]\hline\\[-6pt]
Field of view (FWHP) \dotfill & $60''$ \\
                              & 2.7 kpc at $D$ = 9.3 Mpc \\
Bandwidth       \dotfill & 512 MHz \\
Visibility calibrator  \dotfill & 1418+546 \\
Flux density of calibrator \dotfill & $0.5 - 0.7$ Jy \\
Array configuration        \dotfill & AB, C, and D \\
\hline
\multicolumn{2}{c}{low resolution CO cube} \\
\hline
Weighting \dotfill & Natural, 80 k$\lambda$ taper \\
Synthesized beam   \dotfill &
$4.\hspace{-2pt}''7 \times 3.\hspace{-2pt}''2$, $-54^{\circ}$ \\
                                    & 210 pc $\times$ 140 pc \\
Equivalent $T_{\rm b}$ for 1 Jy beam$^{-1}$ \dotfill & 6.1 K (Jy beam$^{-1}$)$^{
-1}$ \\
Velocity reslution $\Delta v$ \dotfill & 15.6 km s$^{-1}$ \\
rms noise level in channel maps \dotfill & 23 mJy beam$^{-1}$ \\
CO flux within the F.O.V. \dotfill & $340\pm9.1$ Jy km s$^{-1}$ \\
Molecular gas mass within the F.O.V.$^{\dag}$ \dotfill 
& $4.8\times10^8$ $M_\odot$ \\
Peak face-on gas surface density$^{\dag}$ \dotfill 
& $2.0\times10^3$ $M_\odot$ pc$^{-2}$ \\
CO intensity at $15''$ beam$^{\ddag}$ \dotfill & 93.2 K km s$^{-1}$ \\
$I$(NMA)/$I$(45m)$^{\ddag}$ \dotfill & 1.1 \\[4pt]
\hline
\multicolumn{2}{c}{high resolution CO cube} \\
\hline
Weighting \dotfill & Robust (robustness 0) \\
Synthesized beam \dotfill &
\timeform{1".9} $\times$ \timeform{1".8}, $-48^{\circ}$ \\
                                  & 86 pc $\times$ 81 pc \\
Equivalent $T_{\rm b}$ for 1 Jy beam$^{-1}$ \dotfill & 27 K (Jy beam$^{-1}$)$^{-1}$ \\
Velocity reslution $\Delta v$ \dotfill & 15.6 km s$^{-1}$ \\
rms noise level in channel maps \dotfill & 22 mJy beam$^{-1}$ \\
\hline
\end{tabular*}
\vspace{6pt}

\noindent
$^{\dag}$ Adopting a conversion factor $X_{\rm CO} 
= 3.0 \times 10^{20}$ cm$^{-2}$ (K km s$^{-1}$)$^{-1}$.

\noindent
$^{\ddag}$ The NMA CO cube was convolved with $15''$ beam
(the same as the 45 m observations) and
CO intensity at the center (the observed potition with the 45 m),
integrated over the velocity range of $V_{\rm LSR} = 471 - 752$ km s$^{-1}$,
was compared with the 45 m data (table 2).
Most of the single dish flux seems to be recovered in this velocity range.

\end{table*}

\end{document}